\documentclass[twocolumn,prb,aps,amsmath,amssymb,floatfix]{revtex4-1}

\usepackage{graphicx}
\usepackage{epsfig}
\usepackage{amsbsy}

\begin{document}
\title{Simulations of electric-dipole spin resonance for spin-orbit-coupled quantum
dots in Overhauser field: fractional resonances and selection rules}

\author{E.N. Osika, B. Szafran and M.P. Nowak}
 \affiliation{AGH University of Science and Technology, \\ Faculty of Physics and Applied Computer Science,
al. Mickiewicza 30, 30-059 Krak\'ow, Poland}

\date{\today}

\begin{abstract}
We consider spin rotations in single- and two-electron quantum dots that are driven
by external AC electric field with two mechanisms that couple the electron spatial motion
and the spin degree of freedom: the spin-orbit interaction and a random fluctuation of the Overhauser
field due to nuclear spin bath. We perform a systematic numerical simulation of the driven system using a finite difference
approach with an exact account taken for the electron-electron correlation.
The simulation  demonstrates that the electron oscillation in fluctuating nuclear field is translated
into an effective magnetic field during the electron wave packet motion. The effective magnetic field
drives the spin transitions according to the electric-dipole spin resonance mechanism.
We find distinct signatures of selection rules for direct and higher-order transitions in terms of the spin-orbital symmetries of the wave functions. The selection rules are violated by the random fluctuation of the Overhauser field.
\end{abstract}

\maketitle

\section{Introduction}
Control and coherent manipulation of electron spin
in quantum dots\cite{hanson} are extensively studied in the context
of information storage and processing.\cite{lossdivi}
The most basic operation on the single electron spin is its rotation.
One of applicable procedures for spin rotation is the electron spin resonance (ESR). The ESR appears
in constant magnetic field $B$ which splits the spin-up
and spin-down energy levels. In presence of an additional AC magnetic field with frequency tuned to the Zeeman energy splitting,
the electron undergoes the Rabi oscillation between the opposite spin states.\cite{ESR}
It has been proposed for quantum wells \cite{EDSRQW} and quantum dots \cite{EDSR}
that the oscillating external magnetic field can be replaced by oscillating electric field
according to electric-dipole spin resonance (EDSR) mechanism.\cite{EDSRo}
In presence of the spin-orbit (SO) coupling the electron motion induced by the electric field
is translated into an effective magnetic field that depends
on the electron momentum. The effective SO magnetic field
induces Rabi oscillations and the corresponding spin rotations
in a similar manner as the external field.
The EDSR was observed in experiments
that probe  lifting of the Pauli blockade of the current
across double quantum dots.\cite{EDSRm,Sch,Nadj}
It has also been demonstrated that the role of SO coupling that
translates
the  electron motion into appearance of an effective magnetic field
can be played by a gradient of external magnetic field. \cite{Space}
Successful experiments, in which EDSR is achieved due to the spatial
fluctuation of the GaAs nuclear spin field created by the hyperfine interaction (HF)
were also performed.\cite{EDSRnuc}

The purpose of the present paper is the simulation of the EDSR that results
from the random fluctuation of the HF field and determination
of its signatures as compared to the process governed by SO interaction.
We consider single and two-electron systems confined
in quantum dots and transitions that are induced by oscillating electric field.
%We consider
%both mechanisms that couple the electron motion to its spin:
%the fluctuation of the static Overhauser field and the SO interaction.
We perform numerical simulations for systems with strong lateral confinement
using a finite-difference approach that takes an exact
account for the electron-electron correlation.
We discuss the characteristic features of the resonant transitions due to
the HF field and/or SO interaction. In particular the results
indicate existence of selection rules for  transitions governed by  SO interaction.
We discuss the symmetries behind the selection rules also for
fractional resonances that have been observed in
experiments with both SO coupling \cite{Sch,Nadj}
and HF field \cite{EDSRnuc}
and were recently explained\cite{nowak} as second-order processes
similar to multi-photon transitions in quantum optics.\cite{optics}
We demonstrate that the selection rules are violated by the Overhauser field fluctuations.
%The presence of the transitions forbidden by the spin-orbital symmetry could be used as a signature for the presence of the %Overhauser field fluctuation within the quantum dot.
This finding is a relevant information for dynamical nuclear polarization\cite{ot,johnson2005,inni} which reduces the randomness
of the Overhauser field for which restoration of the selection rules should be observed.

\section{Theory}

The EDSR requires application of an external magnetic field to induce
the Zeeman splitting. At non-zero $B$ the  direct electron-nuclear spin flip-flops\cite{johnson2005} become suppressed.\cite{hanson} Then,
the field of nuclear spins separates from the electron wave function.
The characteristic time for the fluctuation of the nuclear spin field
due to the dipole-dipole  and HF interactions
is of the order of 10-100 $\mu$s which is much longer \cite{hanson} than
the time scales for the evolution of the electron spin. For the above reasons
we will treat the HF field as a static Overhauser magnetic field.

\subsection{Hamiltonians}
\begin{figure}[ht!]
\hbox{
	   \epsfxsize=90mm
           \epsfbox {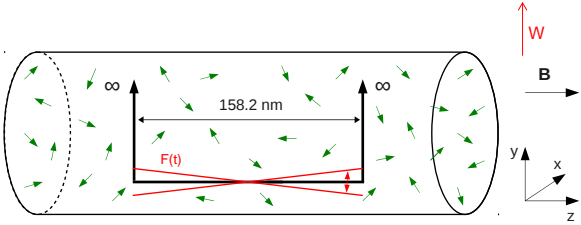}
          }
\caption{ Schematics of the considered system: quantum dot of length $L=158.2$ nm assumed
in form of an infinite quantum well, with strong electron
confinement near the axis of the dot $z=0$, with external magnetic field vector ${\bf B}$ oriented axially,
external electric field perpendicular to the axis ${\bf W}$, and AC electric field along the dot ${\bf F}(t)$.}
\label{SCHEMAT}
\end{figure}

We consider a quasi one-dimensional quantum dot with a strong lateral confinement -- see Fig. \ref{SCHEMAT}
with external magnetic field oriented along the axis of the wire ${\bf B}=(0,0,B)$.
In this configuration the orbital effects of the external magnetic field can be neglected,
and the single-electron Hamiltonian takes the form,
\begin{equation}
h = \frac{\hbar^2 \mathbf{k}^2}{2m^*}+V_c(\mathbf{r}) +V_{AC}(z,t)+\frac{1}{2}g\mu_B B\sigma_{z} + H_{SO} + H_{nuc}
\end{equation}
where $V_c$ is the confinement potential, $V_{AC}=z eF \sin(\omega t)$
is the potential of the AC electric field which will drive the
spin transitions,
$k=-i\nabla$, $H_{nuc}$ describes the interaction of the electron spin with the Overhauser field $\mathbf{B}_{nuc}$
\begin{equation}
H_{nuc} = \frac{1}{2}g\mu_B \boldsymbol{\sigma} \cdot \mathbf{B}_{nuc}({\bf r}),
\end{equation}
and  $H_{SO}$ introduces the linear Rashba spin-orbit interaction due to the electric
field oriented $(0,W_y,0)$ perpendicular to the length of the quantum dot
\begin{equation}
H_{SO} = \alpha(\sigma_z k_x-\sigma_x k_z). \label{cc}
\end{equation}
We apply GaAs material parameters with the effective electron mass  $m^*=0.067m_0$, and the Lande factor $g=-0.44$.
We  assume $W_y=100$ kV/cm which for the bulk SO coupling constant
 $\alpha_b=0.044$ nm$^2$ (after Ref. \onlinecite{silva}) gives
 $\alpha=\alpha_b W_y=0.44$ meV nm. For this value of $\alpha$, SO coupling produces comparable effects to the
Overhauser field fluctuation (see below).

We assume that the confinement potential $V_c$ is separable into a lateral  $V^l$
and longitudinal $V^z$ components: $V_c=V^l(x,y)+V^z(z)$.
We consider a quantum dot with a strongly an elongated geometry (see Fig. \ref{SCHEMAT}).
For the length of the dot which largely exceeds the lateral extent of the
wave function all the low-energy phenomena appear in the ground-state of the lateral quantization.\cite{pt}
We assume that the quantum dot is defined in the $z$ direction by an infinite quantum well of length 158.2 nm: $V^z(z)=V^\infty(z)$  (see Fig. \ref{SCHEMAT}) and that the lateral confinement potential
is strong enough to induce localization around the axis of the cylinder within a region
of radius 10 nm. We assume that the
lateral wave function can be described by a Gaussian for which an analytical form of the effective electron-electron interaction potential is known.\cite{pt}
Namely, we take the single-electron lateral wave function of form
\begin{equation}
 \Psi(x,y) = (\sqrt{\pi} l)^{-1} \exp[-(x^2+y^2)/2l^2], \label{lwf}
\end{equation}
where $l=10$ nm is adopted for the localization parameter.
Upon integration over the lateral degrees of freedom\cite{pt,nowak}
the single-electron Hamiltonian takes a 1D form,
\begin{eqnarray}
\langle\Psi|h|\Psi\rangle&=&h_{1D}=- \frac{\hbar^2}{2m^*} \frac{\partial^2}{\partial z^2} + V_{AC}(z, t)+V^z(z) \\ \nonumber && - \alpha \sigma_{x} k_{z}
+ \frac{1}{2}g \mu_B \boldsymbol{\sigma}\cdot (\mathbf{B}+\mathbf{B}_{nuc}(z)),
\end{eqnarray}
For the electron pair with a double quantum dot is considered in Sec. III. B
we take $V^z(z)=V^\infty(z)+V_b(z)$, where $V_b$ is barrier applied centrally to the system of height  10 meV and width 13.56 nm.

For the electron-pair we consider the Hamiltonian including the electron-electron interaction
\begin{equation}
 H = h(1) + h(2) + \frac{e^2}{4\pi \epsilon_0 \epsilon |\mathbf{r_{12}}|},
\end{equation}
with the  dielectric constant of $\epsilon=12.9$.
After integration of the Hamiltonian with the lateral wave function (\ref{lwf})
one obtains
\begin{equation}
 H_{1D} = h_{1D}(1) + h_{1D}(2) + \frac{\sqrt{\pi/2}}{4\pi \epsilon_0 \epsilon l} \mbox{erfcx}\left[\frac{|z_1-z_2|}{\sqrt{2}l}\right],
\end{equation}
where the last term is the effective 1D interaction derived in Ref. [\onlinecite{pt}]
as $\langle \Psi(x_1,y_1)|\frac{e^2}{4\pi\epsilon\epsilon_0 |{\bf r_{12}}|}|\Psi(x_2,y_2)\rangle $.

Numerical calculations are performed with a finite difference approach with discretized versions
of the differential operators and a finite mesh constant $\Delta z$. For a single electron we look for eigenstates
and time evolution of the spinor components $\Psi=(\Psi_\uparrow(t), \Psi_\downarrow(t))^T$.
For the two-electron wave function we solve the equations for the bi-spinor in form
\begin{equation}
\Psi(z_1,z_2)=\left(\begin{array}{c} \Psi_{\uparrow\uparrow} (z_1,z_2)\\ \Psi_{\uparrow \downarrow } (z_1,z_2)\\ \Psi_{\downarrow \uparrow } (z_1,z_2)\\ \Psi_{\downarrow \downarrow} (z_1,z_2)\end{array}  \right).
\end{equation}
The fermion symmetry with respect to the electron interchange implies that $\Psi_{\uparrow\uparrow} (z_1,z_2)=-\Psi_{\uparrow\uparrow} (z_2,z_1)$ and
$\Psi_{\downarrow\downarrow} (z_1,z_2)=-\Psi_{\downarrow\downarrow} (z_2,z_1)$. The spin unpolarized parts have no definite symmetry with
respect to the exchange of the electron spatial coordinates separately, instead one has
$\Psi_{\uparrow \downarrow } (z_1,z_2)=-\Psi_{\downarrow \uparrow } (z_2,z_1)$.
The evolution of the system in time is given by the Schr\"odinger equation
$i\hbar \frac{d\Psi}{dt}=H\Psi$, which is solved with the explicit Askar-Cakmak scheme\cite{acs}
$\Psi(t+\Delta t)=\Psi(t-\Delta t)+\frac{2 \Delta t}{i\hbar}  H\Psi(t).$

\subsection{Effective magnetic field due to nuclear spins}
\label{ov}

In GaAs each nucleus of the crystal lattice carry an uncompensated spin that
interacts with the electron spin via the Fermi contact hyperfine interaction,\cite{hanson}
$H_{HF}=\sum_k^N A_k {\bf I}_k \cdot {\bf S}\delta({\bf r}-{\bf R}_k)$,
where ${\bf R}_k$ is the position of k-th nucleus, ${\bf I}_k$ and ${\bf S}$ are the nuclear
and electron spin operators, and $A_k$ is the coupling constant which
is proportional to the magnetic moment of nucleus $k$ and
the probability density of finding the electron therein.

\begin{figure}[ht!]
\hbox{
	   \epsfxsize=90mm
           \epsfbox[15 15 670 370] {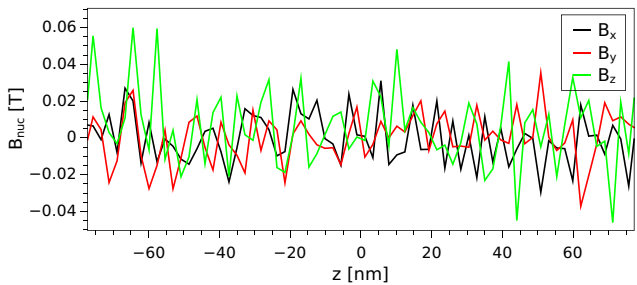}
          }
\caption{Components of the  Overhauser field as integrated with the lateral electron wave
function on a finite difference mesh along the quantum dot (see Section \ref{ov}).}\label{ovf}
\end{figure}

In nonzero $B$ when the entanglement between the electron wave
function and the nuclear spins can be neglected, the electron  interacts
with the ensemble of spins through the Overhauser effective magnetic field ${\bf B_{nuc}}({\bf r})$,
with $H_{HF}\simeq H_{nuc}=\frac{g\mu_B}{\hbar} {\bf B_{nuc}}({\bf r})\cdot {\bf S}$. For fully polarized
nuclei the maximal value of the Overhauser field is ${\bf B_{nuc}}\simeq$ 5T.
For the purpose of the numerical simulation we need the distribution of the Overhauser
field  along the quantum dot. The field is generated in the following manner.
We consider all the nuclei present within the volume of the quantum dot.
With each nucleus at position ${\bf R}_{k}$  (eight nuclei per cubic unit cell
of lattice constant $a=0.565$ nm)
we attribute a local vector of an effective magnetic field
${\bf B}^{\bf r}_k$ of length 5T with orientation taken at random with the uniform distribution.
For the purpose of the present study we do not distinguish between isotopes of Ga and As.
%which have similar properties from the point of view of HF interaction.
The electron lateral wave function $\Psi$ averages the magnetic field
due to the nuclei.\cite{kh,merk}
We calculate the magnetic field in the cell $j$ of the finite
difference mesh, between $z_j$ and $z_j+\Delta z$ points
\begin{equation}{\bf B_{nuc}^j}=
\int_{z_j}^{z_j+\Delta z}dz
\int \int_{-\infty}^{\infty} dxdy |\Psi(x,y)|^2 \sum_k  {\bf B}^{\bf r}_k\delta({\bf r}-{\bf R_{k}}).
\end{equation}
The simulated field due to nuclear spins is plotted in Fig. \ref{ovf}.
For $l=10$ nm, the field amplitude is of the order of 20 mT.
%Since the amplitude of ${\bf B_{nuc}}$ decreases as $1/\sqrt{N}$, where
%$N$ is the number of nuclea seen by the wave function within
%the slice of space $(z_j,z_j+\Delta z)$.
%This implies that the stronger the confinement of the lateral wave function $\Psi(x,y)$ the larger the amplitude of the effective magnetic field.
The actual effective field perceived by the electron spin will
be further reduced to a few mili Tesla\cite{mT} by the spread of the wave function along
the quantum dot (see below).

\section{Results and Discussion}
\begin{figure}[ht!]
\hbox{
	   \epsfxsize=90mm
           \epsfbox[15 0 525 315] {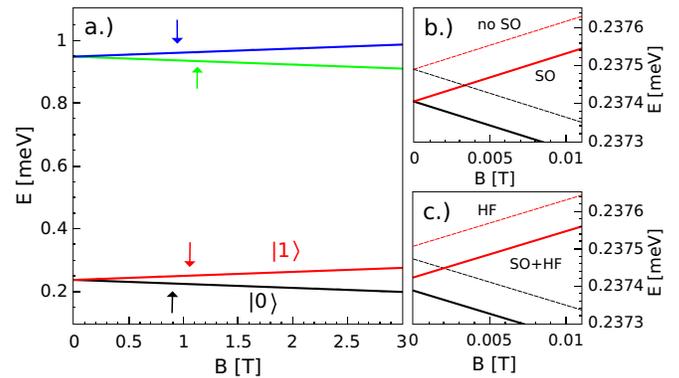}
          }
\caption{(a) Energy spectrum of the single electron in a single quantum dot of length 158.2 nm (see Fig. \ref{SCHEMAT}) as a function of the external magnetic field for
a static potential $F=0$.
The enlarged fragments of weak magnetic fields are given in (b) and (c).
In (b) no HF field is assumed. In (c) we assume the HF field of Fig. \ref{ovf}.
In (b) and (c) the results plotted with solid (dashed) lines were obtained with (without) SO coupling.
}\label{wije}
\end{figure}
\subsection{Single-electron electric-dipole spin resonance}
Figure \ref{wije}(a) shows the single-electron energy spectrum for a quantum dot of length 158.2 nm as a function of the magnetic field.
%and the Zeeman splitting of the energy levels for opposite spin orientations.
Figures \ref{wije}(b) and (c) present  zoom at the ground-state energy level
that is split
by the Zeeman interaction.
The spin-orbit coupling  [Fig. \ref{wije}(b)] lowers the energy
levels and preserves the double (Kramers) degeneracy of the ground-state.
When the Overhauser field  is introduced
[Fig. \ref{wije}(c)], the energy levels are no longer degenerate
at 0T. For the adopted parameters ($l,\alpha$) the energy effects due to both spin-orbit coupling
and hyperfine field are comparable and of the order of  $0.1\mu$eV.
\begin{figure}[ht!]
\hbox{
	   \epsfxsize=90mm
           \epsfbox[15 0 550 308] {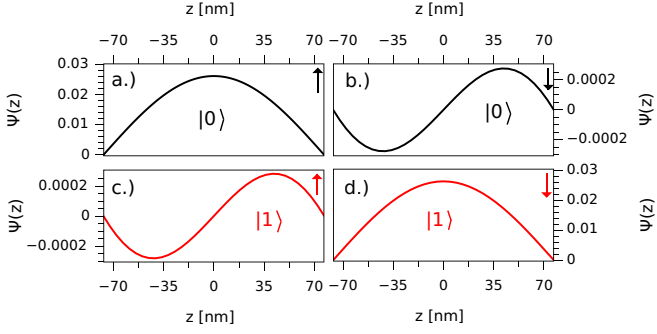}
          }
\caption{Results for a single electron in a single quantum dot of length 158.2 nm (as in Fig. \ref{wije}) with $F=0$ (static potential).  Spin-up (a,c) and spin-down (b,d)  wave functions components of the ground state (a,b) and the first excited
state (c,d) as obtained with SO coupling (no HF field) for the external field of $B=0.3$ T.
The wave functions are given in atomic units.
}\label{falso}
\end{figure}

\begin{figure}[ht!]
\hbox{
	   \epsfxsize=90mm
           \epsfbox[15 0 583 242] {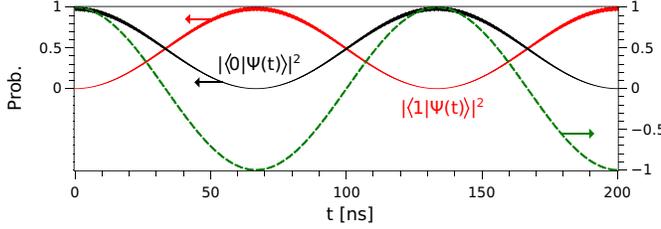}
          }
\caption{
Results for a single electron in a single quantum dot of length 158.2 nm (as in Fig. \ref{wije})
but with the amplitude of AC field $F=0.05$ kV/cm.
EDSR induced by SO coupling (no HF field) for resonant AC frequency
 for the transition
from the ground-state to the first-excited state and $B=0.3$ T oriented
along the $z$ direction.
The black (red) lines show the projections of wave functions on
the two lowest-energy single electron states and the dashed line gives the mean
value of the spin $z$ component that is referred to the right axis. }\label{jeso}
\end{figure}

\subsubsection{EDSR due to SO coupling}
Let us now consider the spin rotations due to spin-orbit coupling
only (no hyperfine field). Figure \ref{falso} shows the real part of the wave functions
of the two lowest energy levels for $B_{e}=0.3$ T.
The ground-state [Fig. \ref{falso}(a,b)] is nearly spin polarized. Its spin-down
component [Fig. \ref{falso}(b)] is of the odd spatial parity
and corresponds to the first-excited energy level of the quantum dot without SO coupling.
The SO coupling Hamiltonian commutes with the s-parity
operator
$P_s=\sigma_z P$, where $P$ is the parity operator
with respect to point inversion through the center of the dot.
Therefore, the components of the spin-orbitals possess a definite but
opposite spatial parities. The SO coupling appears through coupling
of spatial energy levels of opposite parities.

Now, we apply a harmonic AC field of amplitude $F=0.05$ kV/cm
-- for which the potential drop along the dot is $\simeq 0.8$ meV.
We consider the ground-state as the initial condition and
apply the resonant frequency for the transition to the first excited
energy level with inverted spin, $\omega=(E_1-E_0)/\hbar$, which
corresponds to AC oscillation period of about 0.6 ns.
We see in Fig. \ref{jeso} that after 60 ns (nearly 100 periods of
the external electric field) the electron spin gets inverted and the electron occupies the first excited state.

\begin{figure}[ht!]
\hbox{
	   \epsfxsize=90mm
           \epsfbox[15 0 426 571] {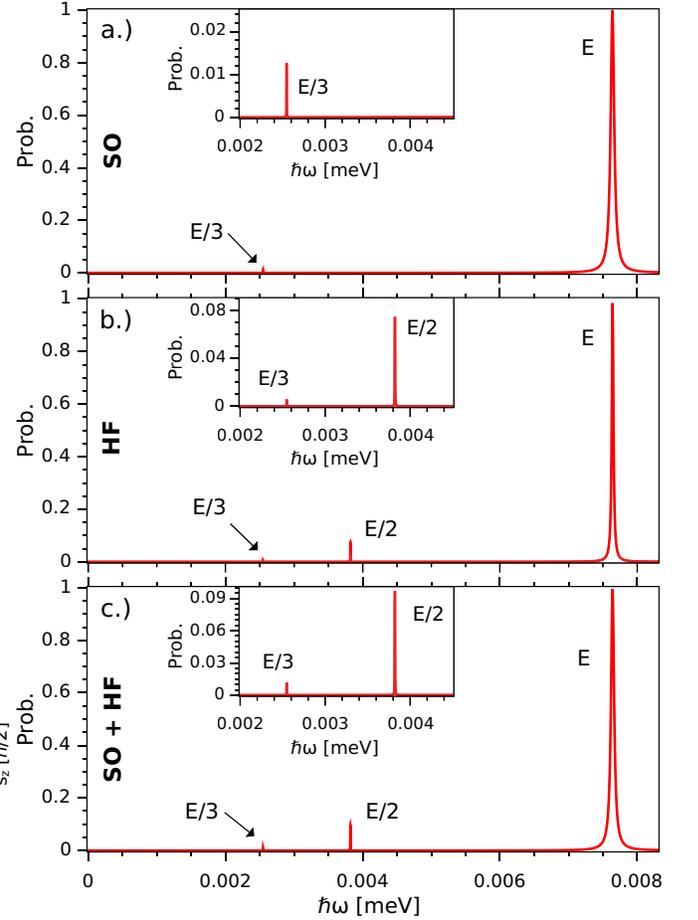}
          }
\caption{Results for a single electron in a single quantum dot of length 158.2 nm with
the amplitude $F=0.05$ kV/cm of the AC field (as in Fig. \ref{jeso}). Maximal probability of the spin inversion as a function of the driving frequency
for 500 ns of the simulation. Separate plots correspond to:
SO interaction without HF field (a), HF field without SO interaction (b),
both HF and SO interaction present (c). Insets show the enlarged fragments
for second and third order transitions.
}\label{skanyje}
\end{figure}

Figure \ref{skanyje}(a) shows the maximal probability of the spin inversion
obtained during 500 ns of the simulation as a function of the frequency
of the AC field. We observe a pronounced resonance near 7.5 $\mu$eV,
that corresponds to the simulation of Fig. \ref{jeso}.
We also notice a sign of fractional $1/3$ resonance near 2.5 $\mu$eV.
Note, that the half-resonance\cite{nowak} due to second-order transition that could be expected near 3.8 $\mu$eV is missing.
The reason for the missing half-resonance is the symmetry of the wave
function for the initial and final states.
The oscillator strength of the direct (first-order) transition from initial state $|i\rangle$ to
the final state $|f\rangle$ which results from the harmonic perturbation of form $z\sin(\omega t)$,
is determined by the Fermi golden rule with the matrix element $\langle i|z|f\rangle =\langle i_\uparrow |z|f_\uparrow\rangle +\langle i_\downarrow |z|f_\downarrow\rangle$.
For the wave functions of Fig. \ref{falso} the components of the sum are non-zero due to the
spin-orbit coupling that introduces non-zero overlap between the components of wave functions for the same
spin orientation. Moreover, since the initial and final states have opposite parities for each of the components,
and $z$ is the odd parity function, the integrands are even functions with respect to point inversion
through the center of the dot. In general, the first-order transition  are allowed between states of opposite s-parities.

According to the time-dependent perturbation theory,\cite{shankar} the second-order transitions  are allowed
if one the products $\langle f|z|m\rangle \langle m|z|i\rangle$ is non-zero, where $m$ is any of
the intermediate eigenstates of the unperturbed Hamiltonian. Now, since all the eigenstates
of the unperturbed Hamiltonian possess a definite s-parity,
and in our case $|i\rangle$ and $|f\rangle$ states are of the opposite s-parity, none of these products
can be non-zero.
The second-order transitions are allowed only between the states of the same s-parity (see below for EDSR in
the two-electron system).
The third order transitions that involve two intermediate states
are again allowed between states of opposite s-parity, hence the 1/3 peak observed in Fig. \ref{skanyje}(a).

\begin{figure*}[ht!]
\hbox{
	   \epsfxsize=150mm
           \epsfbox[15 0 800 400] {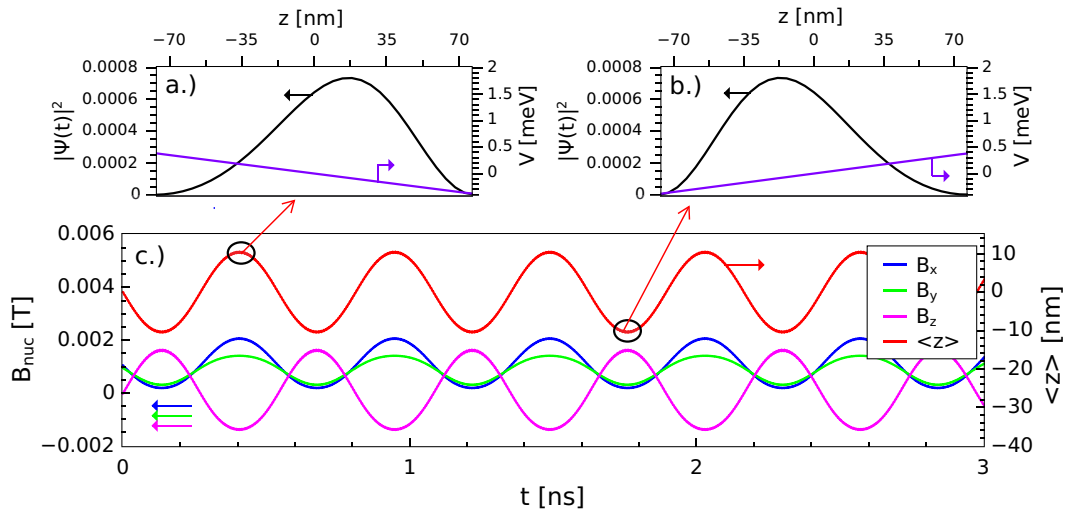}
          }
\caption{Results for a single electron in a single quantum dot of length 158.2 nm with
the amplitude $F=0.05$ kV/cm of the AC field (as in Fig. \ref{jeso}).
(a,b) Electron densities at opposite phases of the driving AC electric field.
(c) The red line shows the mean position of the electron. The three other lines
near the bottom of the plot indicate the average value of the Overhauser field as
perceived by the spin of the electron as a function of time}\label{pn}
\end{figure*}
\subsubsection{EDSR due to the hyperfine field}
%Figure (Rysunek 6) shows the wave functions components that are obtained without
%SO coupling but when the nuclear field is present for the external field of 0.3 T.
%The admixtures of the opposite spin parts to the ground and first
%excited states are of the same form. At a closer inspection one notices
%that the minority spin components are not exactly symmetric functions with
%respect to the inversion through the center of the dot. The deviation is small,
%but relevant for the selection rules.
We now consider the hyperfine field only and apply the harmonic perturbation to the potential (same resonant frequency and amplitude as in Fig. \ref{jeso}).
Charge density at opposite phases of the AC field is given in Fig. \ref{pn}(a,b).
Figure \ref{pn}(c) shows that the center of the electron packet oscillates with an amplitude
of about 10 nm. As the packet oscillates, the electron spin perceives a different
effective magnetic field that is averaged by the moving charge density -- see Fig. \ref{pn}(c)
for the components of the effective field. As a consequence, the electron spin is inverted
after about 200 periods of the oscillating field.

The probability of the spin inversion as a function of $\omega$ are plotted in Fig. \ref{skanyje}(b).
In presence of the nuclear field, the s-parity of eigenstates is no longer defined, and we obtain the direct,
second and third order transitions to the excited state.  A reduced width of the direct transition can be noticed with
respect to the SO case.

\begin{figure}[ht!]
\hbox{
	   \epsfxsize=90mm
           \epsfbox[15 0 550 308] {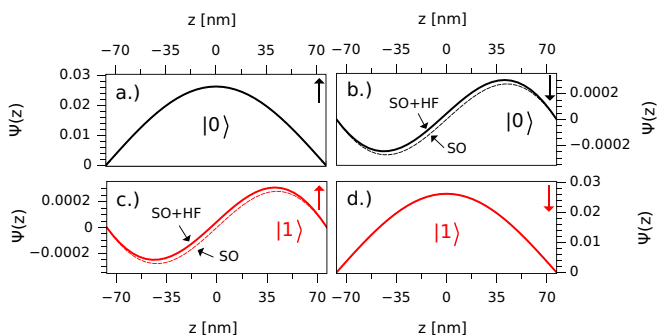}
          }
\caption{
Results for a single electron in a single quantum dot of length 158.2 nm with
the static potential  $F=0$,  as in Fig. \ref{wije}.
Solid lines show the wave function components for the ground-state (a,b) and the first excited
state as obtained with both SO coupling and  HF field present for the external field of $B=0.3$ T.
In the minority spin components (b,c) the results obtained without HF are displayed for comparison.
In the majority spin components (a,d) the lines with and without HF coincide.
}\label{falol}
\end{figure}

\subsubsection{EDSR: combined nuclear field and SO coupling}

Figure \ref{falol} shows the wave functions with both SO coupling and the nuclear field (solid lines).
For comparison, the results without nuclear field are also plotted (dashed lines). The nuclear
field shifts the zero of the minority spin components, which are no longer of a strict odd spatial parity. The impact of the hyperfine field
on the majority wave function is not visible at this scale.
In spite of the fact, that the SO coupling
seems to dominate, the transition probability as function of the frequency of Fig. \ref{skanyje}(c) resembles
rather the case of the nuclear field, due to the presence of the half-resonance. Nevertheless,
the width of the direct transition is similar to the one observed for pure SO coupling.

\subsection{Two-electron results}
\begin{figure}[ht!]
\hbox{
	   \epsfxsize=80mm
           \epsfbox[15 0 525 345] {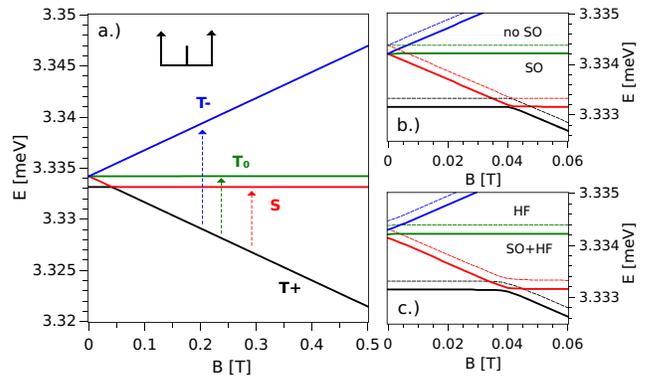}
          }
\caption{(a) Energy spectrum for two electrons in the double quantum dot (within the large
quantum dot of Fig. \ref{SCHEMAT} of length 158.2 nm we put a central barrier
of width 13.56 nm and height 10 meV -- see the inset for the applied
potential) as a function
of the external magnetic field. The static potential is assumed ($F=0$). Arrows indicate the transitions that
appear in EDSR.
The enlarged fragments of weak magnetic fields are given in (b) and (c).
In (b) no HF field is assumed. It is accounted for in (c).
In (b) and (c) the results plotted with solid (dashed) lines
were obtained with (without) SO coupling.
}\label{wde}
\end{figure}

The experiments \cite{EDSRm,EDSRnuc} probe the consequences of EDSR in lifting the Pauli blockade
of the current in double quantum dots. The blockade occurs when electrons in both the dots
acquire the same spin. When the spin of one of electrons is inverted the current restarts to flow.
For the rest of this paper we consider a double
quantum dot [see the inset to Fig. \ref{wde}(a)].
We increased the amplitude of the AC field twice to 0.1 kV/cm -- to maintain
the same potential drop within a single dot as compared to the results of the previous section.
Figure \ref{wde} shows the energy spectrum for the
electron pair. The Coulomb blockade experiments on lifting the Pauli blockade are performed for
the spin-polarized triplet $T+$ as the initial state.

Figure \ref{wde}(b) compares the spectrum without SO and without the hyperfine field (dashed lines),
to the case of pure SO coupling. The avoided-crossing opened by SO interaction between
the $S$ and $T_+$ energy levels is distinctly enhanced by the HF field [see Fig. \ref{wde}(c)],
which also lifts the degeneracy of the triplet states at $B=0$ T.

\begin{figure}[ht!]
\hbox{
	   \epsfxsize=80mm
           \epsfbox[30 10 360 287] {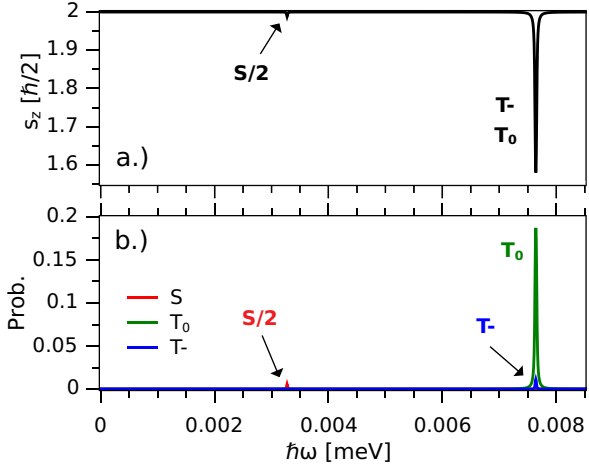}
          }
\caption{Results for two electrons in the double quantum dot considered in Fig. \ref{wde} for the amplitude of AC field $F=0.1$ kV/cm.
Results for SO coupling present but without the Overhauser field. (a) Minimal spin obtained during 100 ns of the simulation
as function of the AC frequency for the $T_+$ as the initial state and $B=0.3$ T.
(b) Maximal probabilities of finding the electron in $S$, $T_0$ and $T_-$ states.
The probabilities are less than 1 due to the finite simulation time.
}\label{deskso}
\end{figure}

\begin{figure}[ht!]
\hbox{
	   \epsfxsize=90mm
\begin{tabular}{c}
           \epsfbox {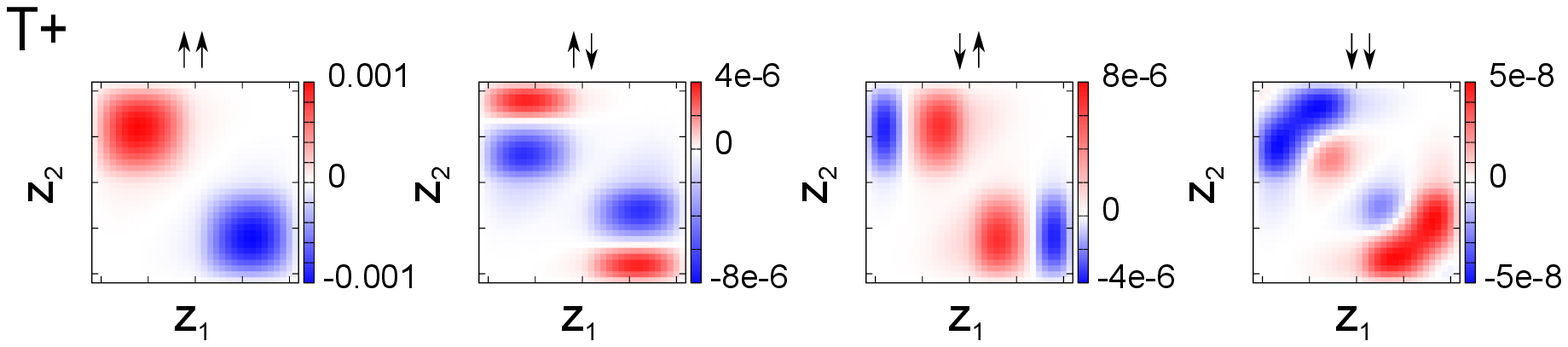} \\
 \epsfbox {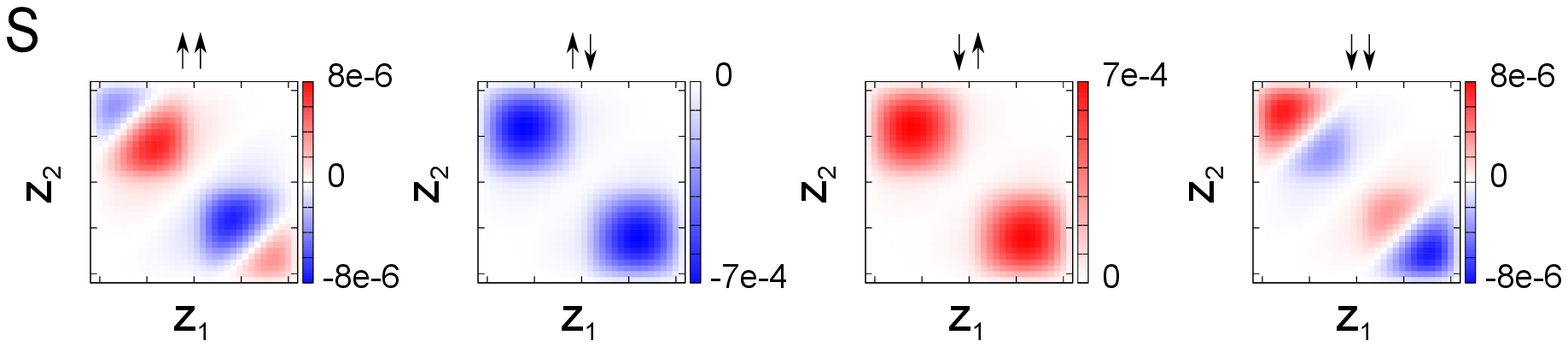} \\
 \epsfbox {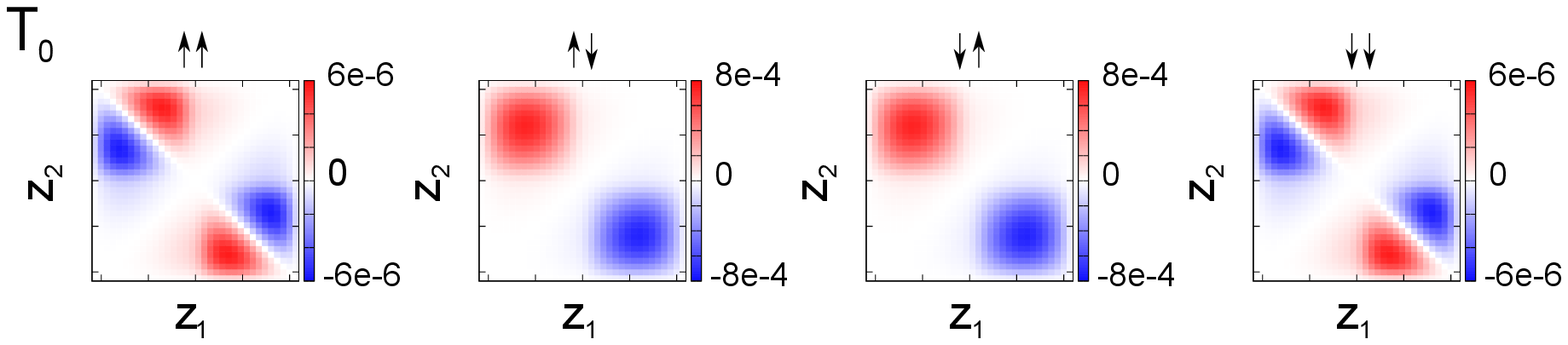} \\

\end{tabular}
          }
\caption{Results for two electrons in the double quantum dot considered in Fig. \ref{wde} for the static potential $F=0$. Components of the two-electron eigenfunctions as functions of both electron coordinates
over the entire length of the double dot for $T_+$ (first row of plots), $S$, and $T_0$
at $B=0.3$ T. The color scales give the real part of the wave function in atomic units.
 Spin-orbit coupling is present, and Overhauser  field  is absent.
The length of $z_1$ and $z_2$ axes corresponds to the length of the double dot $L=158.2$ nm.
}\label{X}
\end{figure}

\begin{figure}[htbp]
\hbox{
	   \epsfxsize=80mm
           \epsfbox[31 311 567 528] {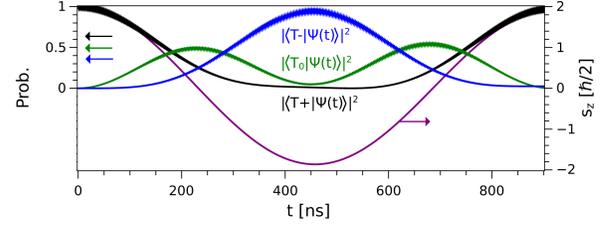}
          }
\caption{Results for two electrons in the double quantum dot considered in Fig. \ref{wde}. Black, green and blue lines show the projections of the wave functions on
the triplets states for resonant frequency tuned to $T_+\rightarrow T_0$ transition
with SO but without HF field. The purple line shows the average value of the $z$ component of the spin.
}\label{czd}
\end{figure}

\begin{figure}[htbp]
\hbox{
	   \epsfxsize=80mm
           \epsfbox[30 10 360 287] {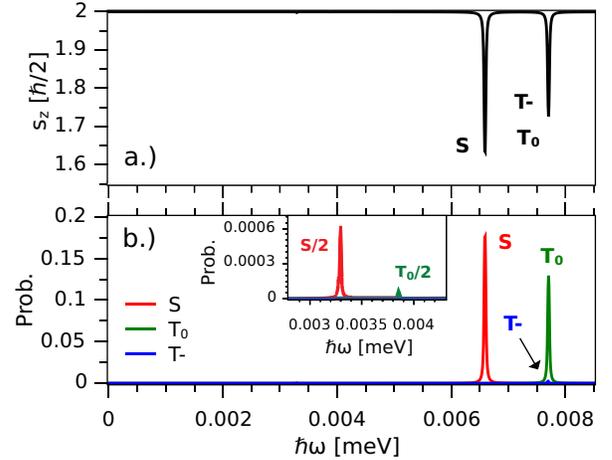}
          }
\caption{Results for two electrons in the double quantum dot [as in Fig. \ref{wde}] for Overhauser field present but without the SO coupling. (a) Minimal spin obtained during 100 ns of the simulation
as function of the AC frequency for the $T_+$ as the initial state and $B=0.3$ T.
(b) Maximal probabilities of finding the electron in $S$, $T_0$ and $T_-$ states.
}\label{skhf}
\end{figure}

\begin{figure*}[htbp]
\hbox{
	   \epsfxsize=160mm
           \epsfbox[6 21 767 318] {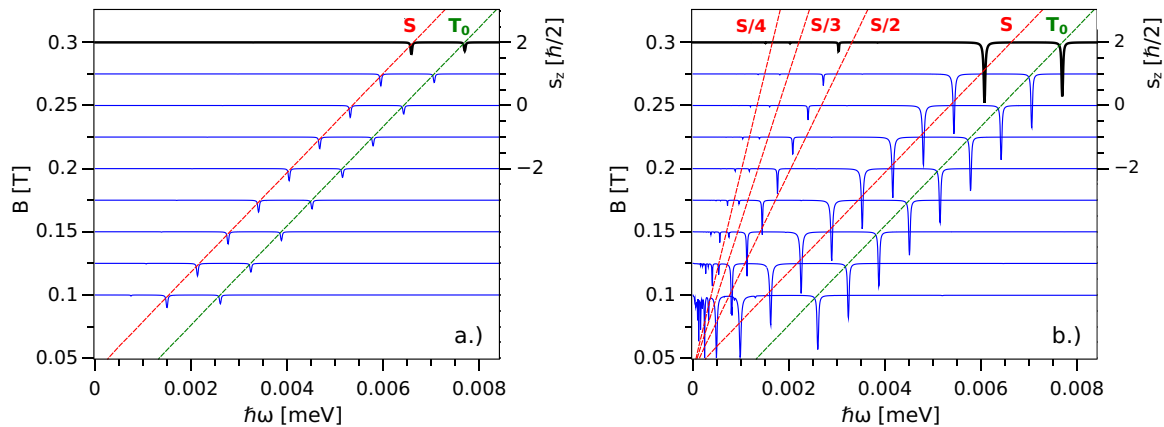}
          }
\caption{Same as Fig. \ref{skhf} only for Overhauser field without SO coupling. Minimal spin obtained during 100 ns of the simulation
for the amplitude of $F=0.1$ kV/cm (a) and $F=0.3$ kV/cm (b) and various values of the magnetic field.
The right scale corresponds to the black curve obtained for external magnetic field of $B=0.3$ T. The blue solid lines
are shifted by 1 down on the spin scale with subsequent values of the magnetic field.
The dashed lines indicate the energies of the transitions from $T_+$ as calculated
from the spectra without perturbation.
}\label{ol}
\end{figure*}

\begin{figure}[htbp]
\hbox{
	   \epsfxsize=80mm
           \epsfbox[30 10 360 287] {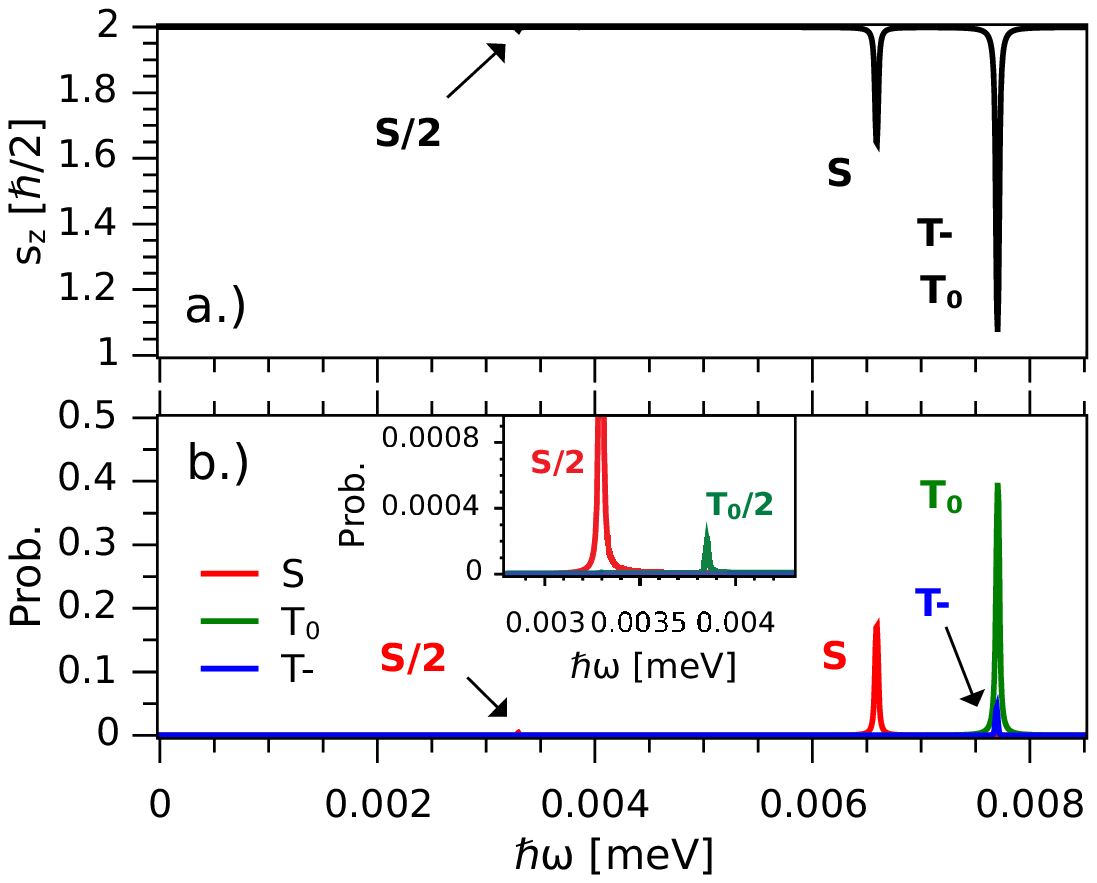}
          }
\caption{Same as Figs. \ref{skhf} only for both Overhauser field and SO coupling present. (a) Minimal spin obtained during 100 ns of the simulation
as function of the AC frequency for the $T_+$ as the initial state and $B=0.3$ T.
(b) Maximal probabilities of finding the electron in $S$, $T_0$ and $T_-$ states.
}\label{skol}
\end{figure}

\subsubsection{Two-electron EDSR: pure SO coupling}
We have performed time dependent simulations starting from the ground state -- the triplet $T_+$ at $B=0.3$T.
The minimal spin that was obtained during 100 ns of the simulation is displayed in Fig. \ref{deskso}(a).
Fig. \ref{deskso}(b) shows the maximal value of the square of projection of the wave function on the Hamiltonian eigenstates.
We can see that the direct transition $T_+\rightarrow S$, that is most relevant for the lifting of the spin Pauli blockade,\cite{EDSRm}
is missing in this plot. The reason for this is again the symmetry of the wave functions,
Fig. \ref{X} shows the real parts of the components of $T_+$, $S$, and $T_0$ wave functions.
Each of the four components possesses a definite parity with respect to simultaneous
inversion of both electron coordinates. In the basis of $(\uparrow,\uparrow)$, $(\uparrow,\downarrow)$, $(\downarrow,\uparrow)$, $(\downarrow,\downarrow)$,
the two-electron s-parity operator has the form
\begin{equation}
P_s=\left(\begin{array}{cccc} P&0&0&0\\0&-P&0&0\\0&0&-P&0 \\ 0&0&0&P \end{array} \right).
\end{equation}
As it can be noticed in Fig. \ref{X}, the only state of positive s-parity is $T_0$. $T_+,S$ and $T_-$ (the last state was not included
in Fig. \ref{X}), have negative s-parity. Hence, the matrix element for the direct transition from $T_+$
to both $S$ and $T_-$ final states vanishes.
On the other hand
the second-order transition to $S$ state with half resonance is observed near $3.2\mu$eV.
No half resonance is observed for the transition to $T_0$, in consistence  with the
discussion of selection rules given above for the single electron.

Figure \ref{deskso}(b) shows that an admixture of $T_-$ state is observed in the final state
for the energy equal half of the energy splitting between $T_+$ and $T_-$.
Note, that in this case the energy spacing between $T_0$
and $T_-$ is exactly the same as between $T_+$ and $T_0$.
Fig. \ref{czd} shows the occupation probability as a function of time for the frequency
of the external field tuned to $T_+\rightarrow T_0$ resonance.
We can see that first a finite probability of $T_0$ occupation appears, but it never approaches
1. Instead the probability occupancy of $T_-$ increases.
Hence in this case we observe a two-step process, first pumping the electron from $T_+$ to $T_0$ state and next from $T_0$ to $T_-$.

\subsubsection{Two-electron EDSR with nuclear field}

Figure \ref{skhf} shows the transitions for the HF field without the SO coupling.
As compared to the previous case we notice: a distinct transition to the $S$ state (s-parities are no longer
definite), and reduced probability of finding $T_-$ in the final state, due to the energy difference
of the transitions between the triplets. Also, a half resonance for transitions to $T_0$ can be observed.
The peaks for half resonances are much smaller than in the SO case. Note, that
also the direct transitions occur significantly slower (cf. the height of the   $T_0$ peak).

In Fig. \ref{ol} we show the scans obtained as a function of both AC frequency
and the magnetic field for AC amplitude of the electric field $0.1$ kV/cm that was considered above (a) and for a stronger amplitude of  $F=0.3$ kV /cm  (b).
In Fig. \ref{ol}(a) at this scale we can see only the direct transitions to $S$ and $T_0$, which exactly agree
in energy with spectral separation of the eigenstates from the ground state (dashed lines).
In Fig. \ref{ol}(b) we notice also fractional transitions
to the singlets. However, there is a detectable redshift of resonant frequencies for singlets.
This shift is due to appearance of the double occupancy of the dot that is induced by stronger
slope of the confinement potential. The double occupancy lowers the energy
of the singlets with respect to the triplets, for which the double occupancy
is forbidden by the Pauli exclusion.\cite{szafranexchange}

The scan of the transitions for both SO and hyperfine field present is given in Fig. \ref{skol}.
As compared to the pure HF case, the height of $S$ peak is reduced with respect to $T_0$.
In contrast to the pure SO coupling case  the half-transitions occur for both $T_0$ and $S$.

\section{Summary and Conclusions}

In summary, we have performed a systematic numerical simulation for the EDSR mechanisms
in single electron (electron spin flip) and two-electron (transitions from the spin-up polarized
triplet to excited states) driven by AC electric field and mediated by both spin-orbit coupling and the abrupt fluctuations
of the static Overhauser field.  We have demonstrated that the latter is translated into a smooth
effective magnetic field felt by the electron spin due to the wave packet oscillation.
The simulation indicates the presence of both integer and fractional resonances that correspond
to first and higher-order transitions. The results for transitions in presence of the pure spin-orbit coupling bear
distinct signatures for the selection rules which result from the spin-orbital symmetry of wave functions.
In presence of the Overhauser field the selection rules no longer hold.
In experiments the randomness of the Overhauser field can be reduced or eliminated by the dynamical nuclear polarization \cite{ot,inni,johnson2005}
or in the strong external magnetic field of a few Tesla. As the hyperfine field becomes ordered, the selection rules should be restored and the observed spectrum for the spin transitions should start to resemble the case of pure SO coupling.

\acknowledgments
This work was supported by the funds of Ministry of
Science and Higher Education (MNiSW) for 2012--2013
under Project No. IP2011038671, and by PL-Grid In-
frastructure. M.P.N. gratefully acknowledges the sup-
port from the Foundation for Polish Science (FNP) un-
der START and MPD programme co-financed by the EU
European Regional Development Fund.

\end{document}